\documentclass{aa}  

\usepackage{natbib}
\usepackage{graphicx}
\usepackage{multirow}
\usepackage{txfonts}
\usepackage{xcolor}

\usepackage[breaklinks=true, colorlinks=true, allcolors=blue]{hyperref}
\begin{document}

   \title{A possible sub-kiloparsec dual AGN buried behind the galaxy curtain}

   \author{P. Severgnini
          \inst{1}
          \and          
          V. Braito
          \inst{1}
          \and          
          C. Cicone
          \inst{2}
          \and          
          P. Saracco
          \inst{1}
          \and
          C. Vignali
          \inst{3}\fnmsep\inst{4}
          \and 
          R. Serafinelli
          \inst{1}
          \and
          R. Della Ceca
          \inst{1}
          \and 
          M. Dotti
          \inst{5}\fnmsep\inst{6}
          \and
          F. Cusano
          \inst{4}
          \and
          D. Paris
          \inst{7}
          \and
          G. Pruto
          \inst{5}
          \and
          A. Zaino
          \inst{8}
          \and
          L. Ballo
          \inst{9}
          \and
          M. Landoni
          \inst{1}}

   \institute{INAF-Osservatorio Astronomico di Brera,
              via Brera 28, I-20121, Milano\\
              \email{paola.severgnini@inaf.it}
             \and
             Institute of Theoretical Astrophysics, University of Oslo, 
             PO Box 1029, Blindern 0315, Oslo, Norway
             \and
             Dipartimento di Fisica e Astronomia, 
             Alma Mater Studiorum, Universit\`a degli Studi di Bologna, 
             Via Gobetti 93/2, I-40129 Bologna, Italy
             \and
             INAF-Osservatorio di Astrofisica e Scienza delle Spazio di Bologna,
             OAS, via Gobetti 93/3, I-40129 Bologna, Italy
             \and
             Dipartimento di Fisica ``G. Occhialini'', Universi\`a di Milano-Bicocca, Piazza della Scienza 3, 20126, Milano, Italy
             \and
             INFN-Sezione di Milano-Bicocca, Piazza della Scienza 3, 20126, Milano, Italy
             \and
             INAF-Osservatorio Astronomico di Roma, Via Frascati 33, I-00078 Roma, Italy
             \and
             Dipartimento di Matematica e Fisica, Universit\`a degli Studi Roma Tre, 
             Via della Vasca Navale 84, I-00146, Roma, Italy
             \and
             XMM-Newton Science Operations Centre, ESAC/ESA, PO Box 78, 28691, Villanueva de la Ca\~nada, Madrid, Spain
             }

   \date{Received 2 October 2020; Accepted 16 December 2020}
 
  \abstract
{
Although thousands of galaxy mergers are
known, only a handful of sub-kiloparsec-scale supermassive black hole (SMBH) pairs have been confirmed so far, leaving a huge gap between the observed and predicted numbers of such objects.
In this work, we present a detailed analysis of the Sloan Digital Sky Survey optical spectrum and of near-infrared (NIR) diffraction limited imaging of \object{SDSS~J1431+4358}. This object is a local radio-quiet type 2 active galactic nucleus (AGN) previously selected as a double AGN candidate on the basis of the double-peaked [OIII] emission line. The NIR adaptive optics-assisted observations were obtained at the Large Binocular Telescope with the LUCI+FLAO camera. We found that   most of the prominent optical emission lines are characterized by a double-peaked profile, mainly produced by AGN photoionization. Our spectroscopical analysis disfavors the hypothesis that the double-peaked emission lines in the source are the signatures of outflow kinematics,
leaving open the possibility that we are detecting either the rotation of a single narrow-line region or the presence of two SMBHs orbiting around a common central potential.
The latter scenario is further supported by the high-spatial resolution NIR imaging: after subtracting the dominant contribution of the stellar bulge component in the host galaxy, we detect two faint nuclear sources at $r<0.5$~kpc projected separation.  Interestingly, the two sources have a position angle consistent with that defined by the two regions where the [OIII] double peaks most likely originate. Aside from the discovery of a promising sub-kiloparsec scale dual AGN, our analysis shows the importance of an appropriate host galaxy subtraction in order to achieve a reliable estimate of the incidence of dual  AGNs at small projected separations.
}


   \maketitle
%

\section{Introduction}
\label{intro}

In the last few  decades, the presence of supermassive black holes (SMBHs) at the centers of massive galaxies and the existence of
correlations between their masses and their host galaxy bulges were observationally uncovered by different authors  \citep[e.g.,][]{sol82,hae93,kor95, fab99,odo02,har04,fer05,kor13,sav16}. 
As a result of gas accretion onto the SMBH, a large amount of energy can be released,  leading to the ignition of an active galactic nucleus \citep[AGN;][]{sol82,hop07,ued14}. 
Galaxy interactions are one of the main processes that can drive gas inflows toward the nuclear region, thus triggering the AGN activity \citep{her89,bar91,mih96,bar02,spr05a,li07,hop09,cape17,blu18}. Therefore, if SMBHs are ubiquitously present in the center of galaxies, the formation of dual AGNs (kiloparsec separation) is expected in a large fraction of merging and merged galaxies.
Following the $\Lambda$ Cold Dark Matter ($\Lambda$CDM) cosmological paradigm predictions, in which galaxies grow hierarchically through (minor and major) mergers, dual AGNs should be 
located in a large number of sources.

Many  of the systematic searches conducted so far are based on the quest for double-peaked narrow optical emission lines originating from two distinct narrow-line regions (NLRs). These studies yield a detection rate of a few
percent of confirmed dual AGNs \citep[see, e.g.,][]{she11,fu12,liu11}. 
The incidence of confirmed dual AGNs in
merging system samples \citep[][]{kos12,ten12,mez14} or among the candidates selected by radio surveys \citep{bur11,fu15a,fu15b,bur18} is slightly higher, but still below 10\%, hence very low with respect to the expectations based on the merger rate of galaxies \citep[e.g.,][]{spr05b,hop05,van12,cape17_b}.
However, it is important to note that all these observational studies suffer from biases that could significantly 
reduce the detection rate of dual AGNs. Some of these biases (e.g., the wavelength used, the incompleteness of the samples, and the difficulty in resolving the pairs at small projected separations) are widely discussed in \cite{sol19} and some of these biases are also addressed in this work.

In principle, if we could assess from observations the exact incidence of dual AGNs, we would  constrain galaxy evolutionary models and study the final stages of the merging process
\citep{van12,ble13,col14}. Moreover, during the formation of
dual systems, significant negative or positive feedback can be produced \citep[see, e.g.,][]{kos12,ble13,mez14,rub20}, affecting both the interstellar and circum-galactic media. Furthermore, dual AGNs are the direct precursors
of binary SMBHs (sub-parsec and parsec-scale separations),  which are amongst the loudest emitters of  gravitational waves in the frequency ranges detectable by pulsar time arrays \citep[PTAs;][]{ver16} and by the future space-based gravitational wave Laser Interferometer Space Antenna \citep[LISA;][]{ama17}.
For these compact systems direct imaging  is beyond the capability (i.e., angular resolution) of any current instrument, except for high-resolution interferometric observations of radio emitting sources \cite[see, e.g.,][but see also \citealt{wro14}]{rod06,kha17,dea14}. 
Regarding the binary SMBHs with projected separation below the very long baseline interferometry (VLBI) resolution limits, an exemplary case is the blazar OJ 287.
For this object a strong indication of the SMBH binary nature comes from  a complex modeling of optical and radio variability \citep{val08,val13,bri18,dey19}.
A still scarce list of few additional promising binary SMBH candidates is available in the literature \citep[see, e.g.,][]{bor09,tsa11,era12,dec13,ju13,she13,liu14,gra15,run17,wan17,sev18,ser20}.

Although  observationally more accessible, even dual AGNs at kiloparsec and sub-kiloparsec  separations require stringent angular resolution to be directly imaged, and 
the current instrumental  limitations make their detection 
particularly challenging.  
Among the few tens of SMBHs at $<$10 kpc separation confirmed so far,  only a few ($\sim$10) are unambiguous sub-kiloparsec pairs of SMBHs \citep[see][and references therein for exhaustive lists of confirmed dual AGNs]{das18,rub18,der19}. 
At present, about 30 percent of the confirmed dual AGNs have been selected from double-peaked 
narrow optical emission line sources \citep{fu11a, mcg15}.   
In this respect, the catalogs of \cite{wan09}, \cite{smi10}, \cite{liu10}, and \cite{ge12}
provide more than a thousand of unique AGNs with a double-peaked [OIII]$\lambda$5008$\AA$ line
identified using the Sloan Digital Sky Survey
Data Release 7 \citep[SDSS-DR7;][]{aba09}. We note that, for 
low-redshift sources (i.e., $z\leq$0.15), the 3" diameter of the SDSS
data fiber ensures that the double-peaked emission lines come from a physical region smaller than 10 kpc.
Unfortunately, the  SDSS fiber does not allow us to discern whether the double-peaked emission lines are due to dual  nuclei or  to kinematical  effects occurring in the surrounding of a single AGN, that is,  jet-cloud interactions or outflows \citep{hec84,whi05,ros10,gab17}, a rotating, disk-like NLR \citep{xu09,mul11}, or the combination of a blobby NLR and extinction effects \citep{cre10}. 
Therefore, follow-up observations, including high-spatial resolution imaging  combined with spatially resolved spectroscopy, have been used by different authors to further investigate
the kinematical origin of double-peaked emission lines \citep[see, e.g.,][]{fu11a,fu12,mcg15,tin11,com12,mul15}. 

In principle, high-spatial resolution (sub-arcsec) observations allow us to identify dual AGNs down to sub-kiloparsec scale. Nevertheless, a strong limitation is that strict observability constraints are often needed to achieve the required sub-arcsec imaging quality, which limits the number of dual AGN candidates that could be investigated in detail so far.
This is particularly true for radio-quiet objects, which represent the majority of the AGN population. 
Optical and near-infrared (NIR) imaging  performed with adaptive optics (AO) systems requires both optimal seeing conditions and  the presence of at least one bright point-like source within much less than one arc minute by the target. 
X-ray observations could reach the sub-arcsec spatial resolution
with the {\it Chandra} sub-pixel imaging technique \citep[see, e.g.,][]{fab18}. However, in this case, a rather large amount of net counts
(at least 500-1000) are required, limiting the studies to very few bright nearby dual AGN candidates.

By combining the results of the four spectroscopic samples quoted above \citep{wan09,smi10,liu10,ge12}, there are 
$\sim$930 AGNs at $z\leq$0.15 with a double-peaked [OIII]$\lambda$5008$\AA$ line and  only 5\%  of these objects have been followed-up with sub-arcsec spatial resolution imaging in the optical and/or NIR bands \cite[i.e.,][]{fu12,mcg15}. For only a small fraction ($\sim$17\%), a companion within 3" (less than 10 kpc of projected separation at z$\leq$0.15) has been directly discerned in the sub-arcsec resolution imaging and thus further investigated through spatially resolved spectroscopy as possible dual AGNs.
As we discuss in this paper, the fraction of dual AGNs on kiloparsec and sub-kiloparsec scales could be higher than that inferred by searching for a companion directly visible on sub-arcsec imaging. 
Even high-resolution imaging may miss the presence of multiple nuclear components if their emission is dimmed by dust and gas and/or blended with a dominant diffuse stellar bulge component from the host galaxy.

In this paper we present the results obtained on a  radio-quiet\footnote{The source is undetected down to about 1 mJy (5$\sigma$) at 1.4 GHz in the VLA FIRST survey \citep{bec95}.} dual AGN candidate: \object{SDSS~J143132.84+435807.20}  (type~2, z=0.09604, hereafter \object{SDSS~J1431+4358}), which belongs to the samples of  \cite{wan09} and \cite{ge12}.
The NIR follow-up was performed with
 the LUCI camera, in diffraction-limited mode, at the Large Binocular Telescope (LBT).
As we show in this work,
to maximize the possibility of identifying multiple nuclei, the dominant stellar contribution should be adequately taken into account and subtracted.

This galaxy shows a clear double-peaked [OIII] emission line  in the SDSS--DR7 optical spectrum  \citep{wan09,ge12}, which prompted a follow up by \cite{nev16} with spatially resolved
long-slit spectroscopy.
In Section~\ref{target}, we summarize the previous results, while in Section~\ref{obs} we present our analysis of the SDSS--DR15{\footnote{The SDSS--DR15 acronym means that the spectrum was downloaded by the Sloan Digital Sky Survey Data Release 15  website: http://skyserver.sdss.org/dr15/en/home.aspx}} spectrum, along with the observations and the analysis we performed on the LBT and on the {\it Swift}-X-ray Telescope 
({\it Swift}-XRT) data of \object{SDSS~J1431+4358}. The results are presented in Section~\ref{results}. In Section~\ref{discussion}, we combine the
spectroscopical information with the NIR results in light of the most plausible physical scenarios for
\object{SDSS~J1431+4358}. Section~\ref{summary} presents our conclusions.

Throughout the paper we assume a flat $\Lambda$CDM cosmology with H$_0$=69.6 km s$^{-1}$ Mpc$^{-1}$ , $\Omega_{\lambda}$=0.7 and $\Omega_{M}$=0.3. All the magnitudes are in the Vega system.
Errors are given at 68 percent confidence level.

\section{SDSS~J1431+4358}
\label{target}

\object{SDSS~J1431+4358}  is a local (z=0.09604) bulge-dominated galaxy (see also Sect.~\ref{lbt}) which, from a visual inspection, appears isolated and  undisturbed \citep[see also][]{ge12}. 
It belongs to a sample of 87 SDSS--DR7 type 2 AGNs with double-peaked [OIII] profiles selected and analyzed by Wang et al. (2009).
In that work, the [OIII] line
was fit with a blueshifted ($\Delta\lambda_{\rm b}$=-2.89$\pm0.33\AA$) and a redshifted ($\Delta \lambda_{\rm r}$=3.23$\pm0.35\AA$) component, neither of which is at the systemic velocity of the host galaxy. The corresponding fluxes of the blueshifted and redshifted components as estimated by Wang et al. (2009) are as follows: $F_{\rm{[OIII]blue}}$=(66$\pm$9) $\times$10$^{-17}$ erg cm$^{-2}$ s$^{-1}$ and  $F_{\rm{[OIII]red}}$=(60$\pm$9)$\times$10$^{-17}$ erg cm$^{-2}$ s$^{-1}$. 
The source is also included in the double-peaked emission-line galaxy sample selected by \cite{ge12}. These authors investigated the nature of each peak component through the 
Baldwin-Phillips-Terlevich diagram \citep[BPT,][]{bal81}  and inferred that both the blue and red components for \object{SDSS~J1431+4358} are produced by AGN photoionization.

\object{SDSS~J1431+4358} was further observed by
\cite{nev16}  with the Apache Point Observatory Dual Imaging Spectrograph (0.42$\arcsec$/pixel in the blue channel, 0.4$\arcsec$/pixel in the red channel). 
The authors investigated the origin of double-peaked narrow lines for  a sample of double-peaked type 2 AGNs at z<0.1 through spatially resolved long-slit spectroscopy.
Observations at two position angles were obtained for each source to constrain the orientation of the NLR. 
 As quoted by \cite{nev16}, while they  were unable to directly identify dual AGNs with long-slit data alone, they defined a classification method 
 to distinguish NLR kinematic classification as rotational or outflow-dominated. 
 As for \object{SDSS~J1431+4358}, the  two  positional angles used during the 
 observations  were orthogonal ($PA_{\rm obs1}$=87 $\deg$ and $PA_{\rm obs2}$=177 $\deg$) and one of these is  
 coincident with  the photometric major axis of the host galaxy, as derived from the SDSS r-band photometry ($PA_{\rm gal}$=87 $\deg$).
 Following  \cite{nev16}, the inferred NLR position angle is $PA_{\rm [OIII]}$=59 $\deg$.
By analyzing the double-peaked profiles at each spatial position, the authors found the possible 
presence of a broad [OIII] emission line component  ($\sigma>$500 km s$^{-1}$)  in the spectrum collected along the positional angle of 
 $PA_{\rm obs1}$=87 $\deg$.
 Although the very large uncertainties associated with the velocity dispersion value ($\sigma$=578$\pm$283 km s$^{-1}$)  make the 
 detection of a broad [OIII]  component only tentative,   \cite{nev16} classified the NLR kinematics as possibly outflow-dominated  \citep[see also ][]{mul11}.
 
As discussed in the next sections, although the spectral analysis does not rule out the presence of an ionized outflow component, 
we find that this mechanism can hardly be the main origin of the double-peaked emission lines observed in the optical spectrum of \object{SDSS~J1431+4358}.

\begin{figure}
   \centering
        \includegraphics[width=\hsize]{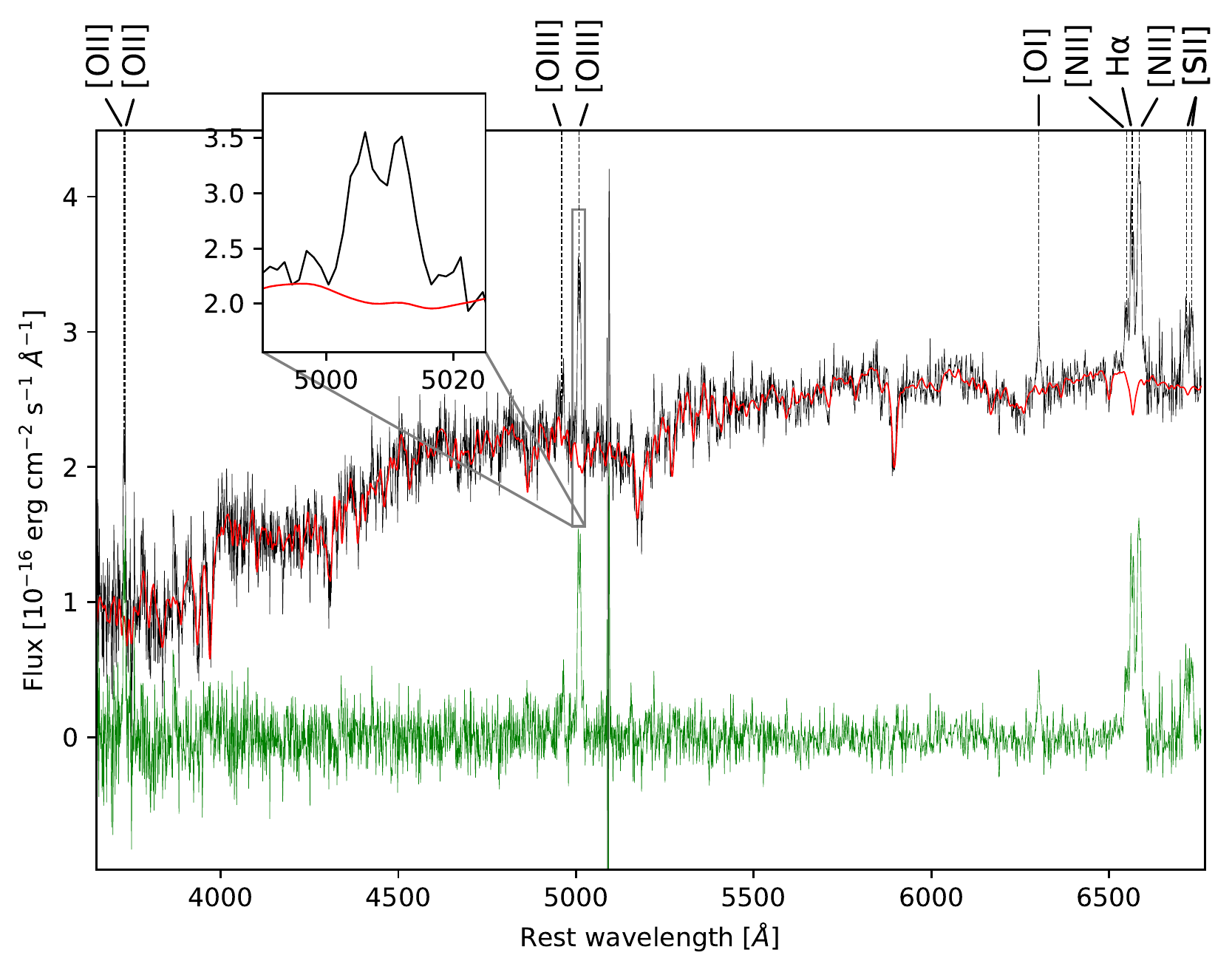}
    \caption{\object{SDSS~J143132.84+435807.20} rest-frame SDSS--DR15 spectrum (black line). The strongest emission lines are labeled and an inset of the spectrum covering the [OIII]$\lambda 5008 \AA$ (rest--frame vacuum wavelengths) emission line is shown.  The red curve indicates the best-fitting MILES model resulting from the pPXF spectral fitting (see Sect.~\ref{optical}), while the green spectrum
    indicates the residual obtained after the subtraction of the stellar component from the observed spectrum.}
\label{sdss15_spectrum}
\end{figure}

\section{Observations and data analysis}
\label{obs}

\subsection{The SDSS--DR15 optical spectrum}
\label{optical}

We analyzed the publicly available SDSS-DR15 optical spectrum of \object{SDSS~J1431+4358} (see Fig.~\ref{sdss15_spectrum})  focusing on the presence and the properties of double-peaked emission line profiles.
We first fitted the host galaxy spectral stellar continuum using the penalized PiXel-Fitting method \citep[pPXF; ][]{cap04,cap17} with the Medium resolution INT Library of Empirical Spectra (MILES) single stellar population (SSP) models \citep[][]{vaz10}. We did not allow pPXF to fit the nebular emission lines as well, since we do this fit in a second stage using our own procedures.
The obscuration of the active nucleus allowed us to determine the redshift 
of the host galaxy ($z$=0.09597$\pm$(6$\times$10$^{-5}$)) through the stellar absorption lines.
In  Fig. \ref{sdss15_spectrum} the best-fitting MILES model is overlaid on the rest-frame SDSS-DR15 spectrum. The residuals obtained by the subtraction of the stellar component from the observed spectrum
are also shown. 
In the following, we present the analysis performed on the residual spectrum.
We verified that our results 
do not depend (within the 1$\sigma$ uncertainties) on the subtracted stellar continuum by running the spectral fitting subtraction several times and with different templates.

For the spectral modeling we used the Python package SHERPA{\footnote{https://sherpa.readthedocs.io/en/latest/}}.
To investigate  the  presence and the properties of the double-peaked profiles, we mainly focused on the [OIII]$\lambda$4960\AA$,\lambda$5008\AA, H$\alpha$ + [NII]$\lambda$6550\AA$,\lambda$6585\AA,~and [SII]$\lambda$6718\AA$,\lambda$6733\AA~emission lines.
For the remaining and less prominent emission lines visible in the spectrum (i.e., [OII] and [OI]; see Fig.~\ref{sdss15_spectrum}) the spectral fitting analysis does not allow us to reach reliable conclusions. In particular, the [OII] emission line is a doublet ($\lambda$3727\AA~and $\lambda$3730\AA);  these emission lines are too close in wavelengths and would require
a higher spectral quality  (both in term of spectral resolution and signal-to-noise ratio) to detect and properly model any double peak. Regarding the [OI] emission line, it could be equally well fitted by two narrow Gaussian components or by a single one plus broad wings. A broad component is often observed for the [OI], which is generally interpreted as evidence that the emitting clouds are optically thick to the ionizing radiation \citep{ost91}.
In the  observed spectrum the H$\beta$ line is only seen in absorption. 
However, once we subtract the dominant galaxy contribution, as shown in Fig.~\ref{sdss15_spectrum}, a  weak  narrow emission component emerges.   Owing to the low signal-to-noise ratio 
(S/N$\sim$3 at the emission line peak), it is not possible to statistically distinguish if the H$\beta$ emission line profile is better reproduced by a single or  double emission line components. We performed the spectral fitting of the  residual spectrum in the rest-frame 4820--4900$\AA$ spectral range  by assuming both a single and double Gaussian components.
In the first case, we found that the 
H$\beta$ emission line can be well reproduced by a narrow (full width at half maximum, FWHM$\sim$390 km s$^{-1}$, $F_{\rm{H\beta}}\sim$19 $\times$10$^{-17}$ erg cm$^{-2}$ s$^{-1}$) Gaussian component at the systemic velocity of the host galaxy. The best-fit parameters found by assuming two
Gaussian components are reported in Table~\ref{sdss15_oiii}. In this case, since the widths of the emission components result unconstrained,
we assumed that they are on the same order, in units of km s$^{-1}$, of those estimated for the strongest narrow emission lines present in the spectrum (see below). 
\begin{figure}[h!]
   \centering
        \includegraphics[width=\hsize]{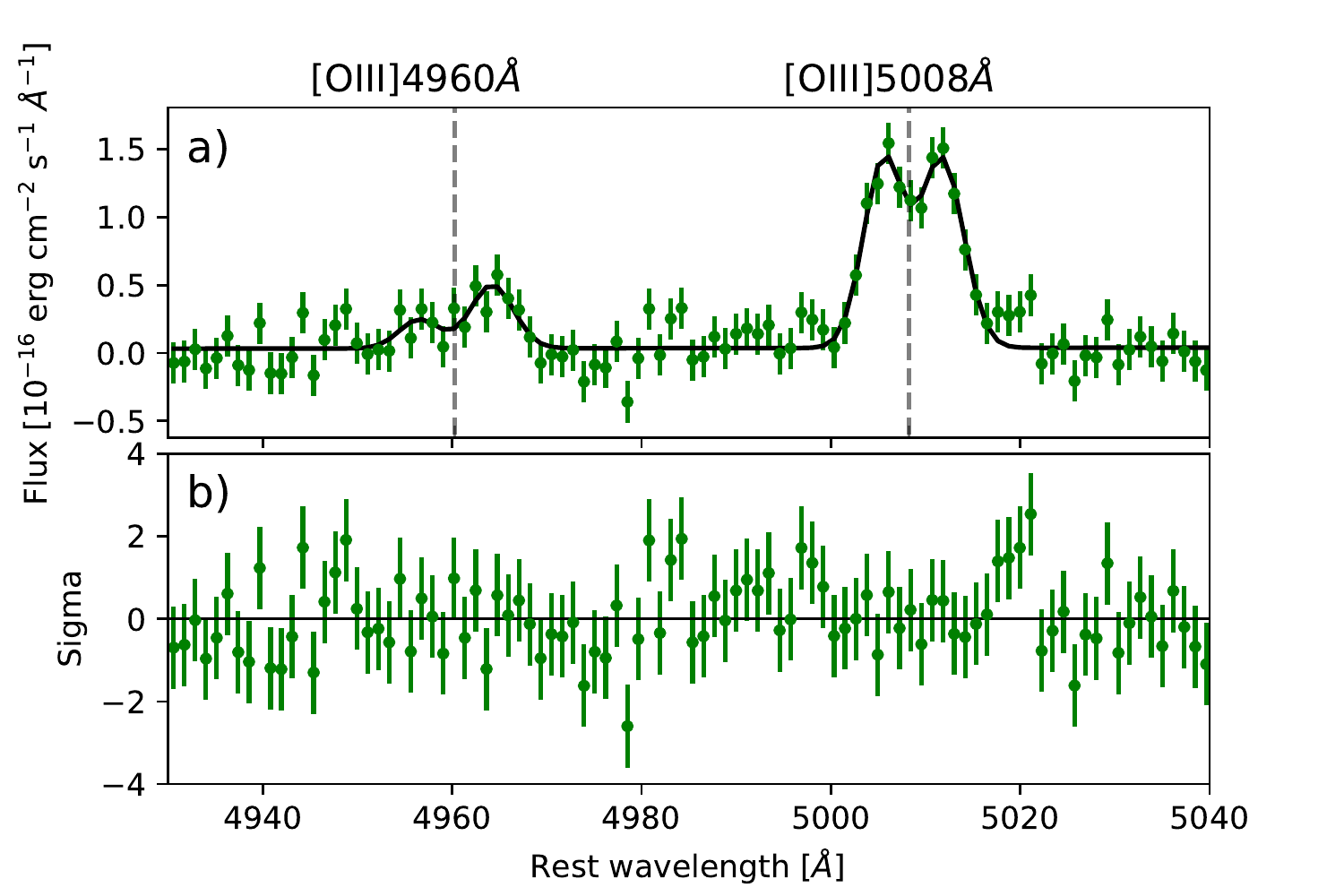}
        \includegraphics[width=\hsize]{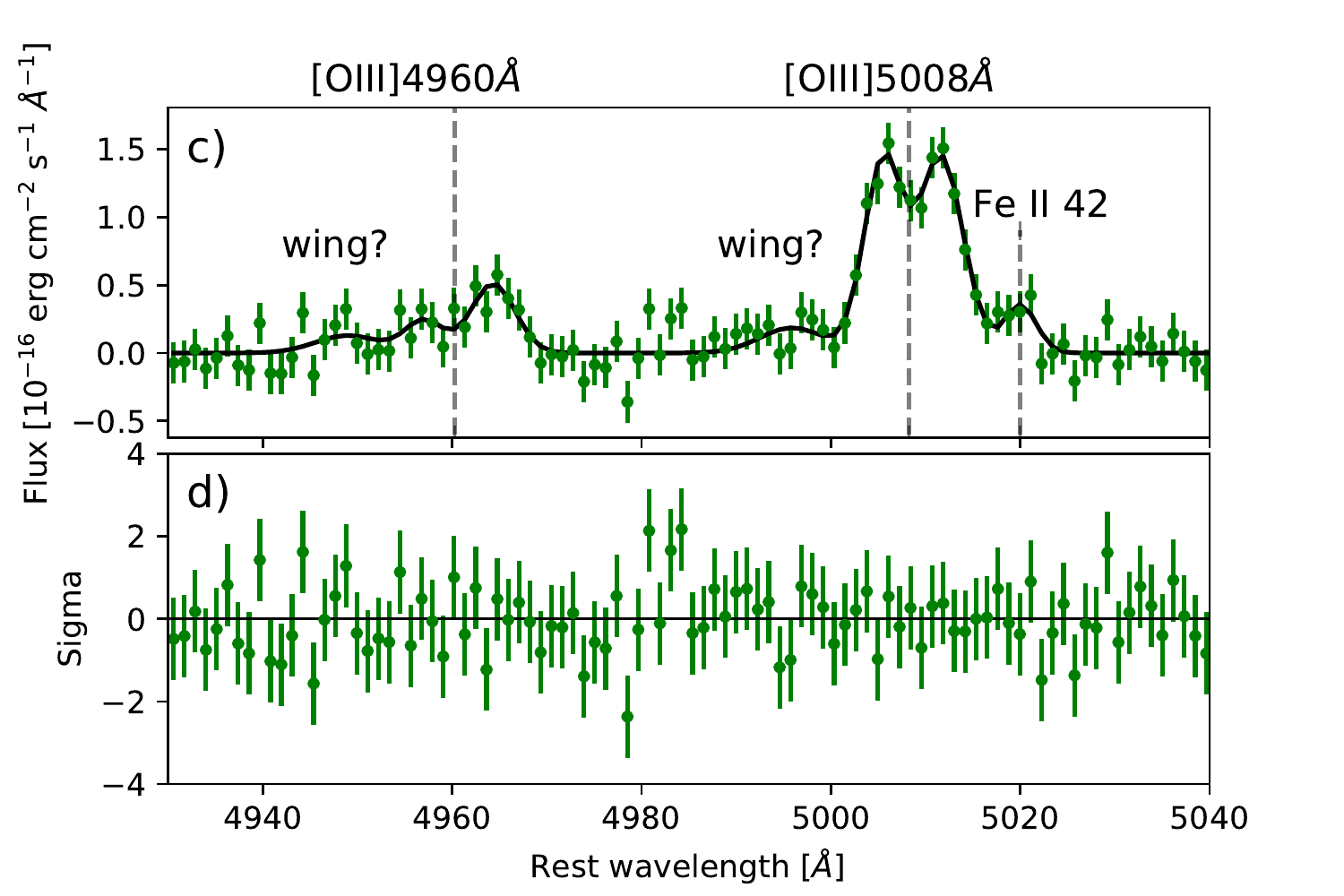}
    \caption{Residual spectrum of \object{SDSS~J1431+4358} (green symbols in this figure and in Fig.~\ref{sdss15_spectrum}) around the  [OIII]$\lambda$4960\AA$,\lambda$5008\AA~ emission line region. The spectrum is shown in the rest-frame vacuum  wavelengths. The gray dashed vertical lines indicate the position of the emission line at the systemic velocity of the host galaxy as derived by the stellar absorption
lines. Panel a): The black solid line represents the model obtained by the combination of a power-law continuum plus two Gaussian emission line components for each of the [OIII] lines. 
Panel b): Relevant residuals, shown in terms of sigmas. Panel c): The black solid line indicates the model obtained by adding a  Gaussian component at 5020\AA~(denoted as Fe II 42) and two broad Gaussian components (denoted as "wing?", see Sect.~\ref{optical}) to the basic model shown
in panel a). 
Panel d): Relevant residuals, shown in terms of sigmas.}
\label{OIII_fit}
\end{figure}

\begin{figure}
   \centering
        \includegraphics[width=\hsize]{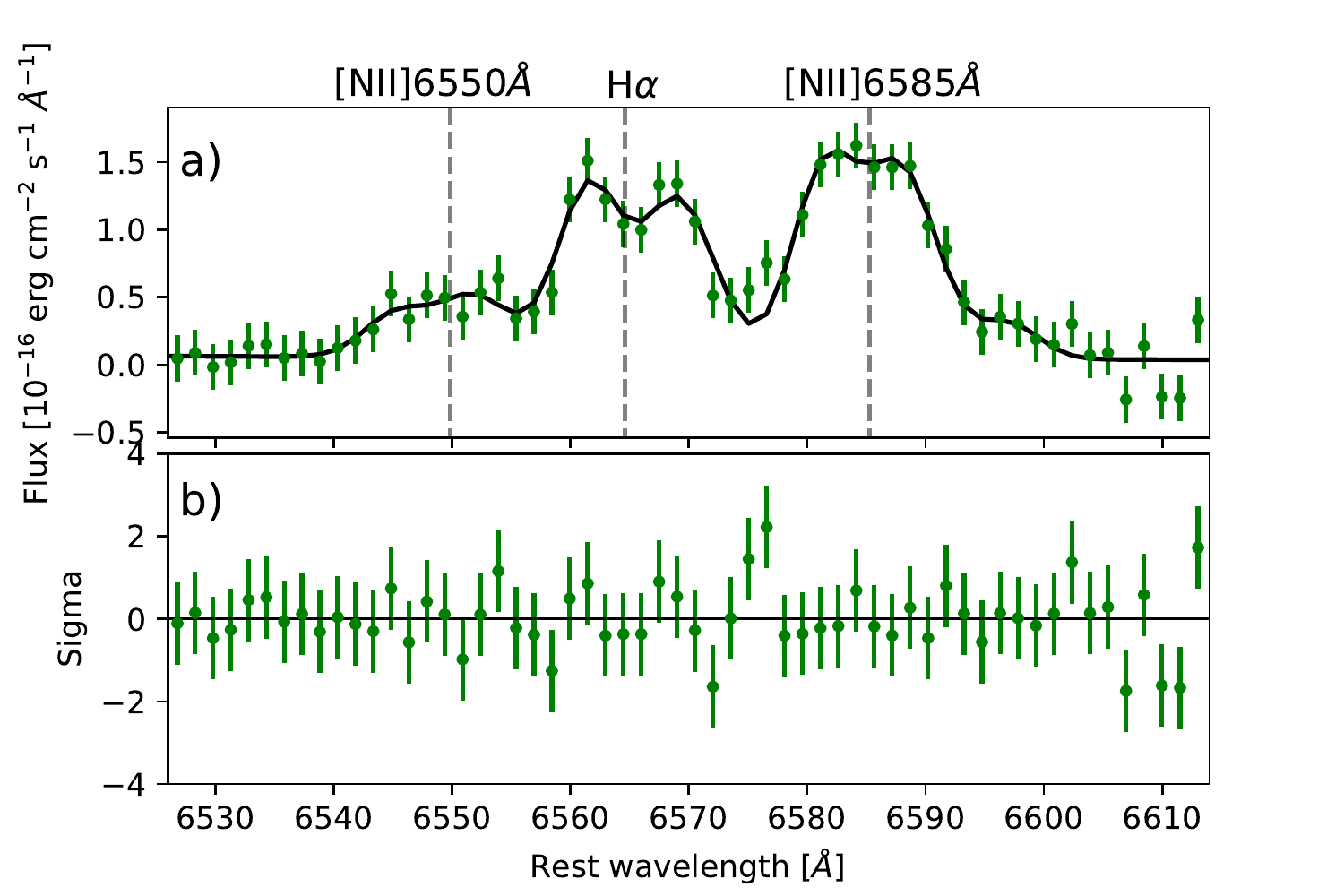}
    \caption{Residual spectrum of \object{SDSS~J1431+4358} (green symbols in this figure and in Fig.~\ref{sdss15_spectrum}) around the  H$\alpha$ + [NII]$\lambda$6550\AA$,\lambda$6585\AA~ emission line region. The spectrum is shown in the rest-frame vacuum  wavelengths. The gray dashed vertical lines mark the position of the emission line at the systemic velocity of the host galaxy. Panel a): the black solid line is the model obtained by the combination of a power-law continuum plus two Gaussian emission line  components for both the H$\alpha$ and for each of the [NII] emission lines. Panel b): relevant residuals, shown in terms of sigmas.}
\label{Halpha_fit}
\end{figure}

\begin{figure}
   \centering
        \includegraphics[width=\hsize]{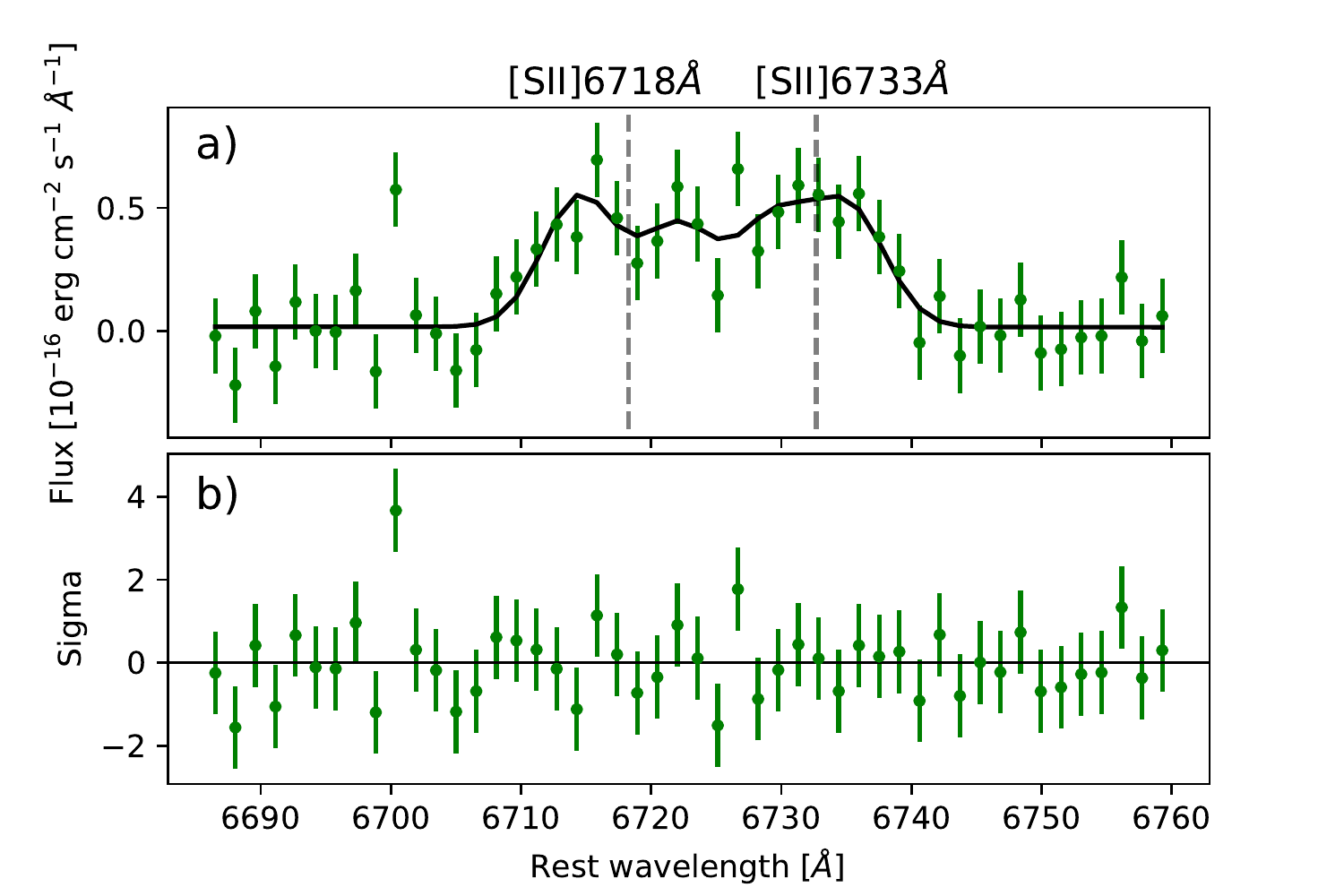}
    \caption{Residual spectrum of \object{SDSS~J1431+4358} (green symbols in this figure and in Fig.~\ref{sdss15_spectrum}) around the  [SII]$\lambda$6718\AA$,\lambda$6733\AA~ emission line region. The spectrum is shown in the rest-frame vacuum  wavelengths. The gray dashed vertical lines indicate the position of the emission line at the systemic velocity of the host galaxy. Panel a): The black solid line represents the model obtained by the combination of a power-law continuum, plus two Gaussian emission line  components for each of the [SII] emission lines. Panel b): Relevant residuals, shown in terms of sigmas.}
\label{SII_fit}
\end{figure}

[OIII]$\lambda$4960\AA$,\lambda$5008\AA~emission lines -- As shown in the zoomed part of the spectrum
in Fig.~\ref{sdss15_spectrum}, the [OIII]$\lambda$5008\AA~ emission line clearly presents a double-peaked profile.
We considered  the residual spectrum (see Fig.~\ref{sdss15_spectrum}) in the rest-frame 4930-5040$\AA$ spectral region.  This covers both the [OIII]$\lambda$4960\AA~and the [OIII]$\lambda$5008\AA~ emission lines and excludes
spectral zones  where absorption lines are present. We used 
a combination of a power-law continuum plus two Gaussian components for each of the [OIII] emission lines. The emission line centroids and their relative intensities were set to be all independent. We allowed their widths to be free to vary, but we kept the width of the blue (red) [OIII]$\lambda$4960\AA~ emission  line component tied to  the blue (red) [OIII]$\lambda$5008\AA~line component. 
The results of the fit are shown in  Fig.~\ref{OIII_fit}, panels a) and b), 
where  the only significant residuals ($\sim$3$\sigma$) are visible around 5018-5020\AA. The
addition of a narrow (FWHM$\sim$250 km s$^{-1}$) Gaussian emission line component at 5020\AA~statistically improves the fit at more than 99.99\% confidence level. 
The wavelength of this narrow line  coincides with the rest-frame wavelength of the strongest Fe~II emission line group emitting between 4930-5050$\AA$ (a6S--z6Po transition, multiplet 42), typically observed in the type 1 and narrow line Seyfert~1 spectra \citep[see e.g][]{ver04}. We checked that  
the other emission lines associated with the same iron transition (i.e., around at 4924\AA~and around at 5169\AA) are also present in the \object{SDSS~J1431+4358} spectrum.
With the exception of the iron excess, only very low statistical significance ($\sim$2$\sigma$) residuals are present in the residual shown in panel b) and no further  components are requested by the fit.
In spite of this, following the \cite{nev16} results on the possible presence of an outflowing material, we also included two additional broad Gaussian emission line components (see panels c) and d) of Fig.~\ref{OIII_fit}). The first component is around the 4940-4955\AA~excess and a second is around the 4990-5000\AA~excess (both denoted as "wing?" in Fig.~\ref{OIII_fit}). For both of these, we found similar FWHM ($\sim$500 km s$^{-1}$) and similar offset central wavelength ($\sim$-700 km/s) with respect to the [OIII]$\lambda$4960\AA~and the [OIII]$\lambda$5008\AA~emission lines, respectively. 

Best-fit values relevant to the [OIII] narrow emission lines are reported in Table~\ref{sdss15_oiii}; we note that they do not depend on the adopted model (with or without broad wings).
Our results on the [OIII]$\lambda$5008\AA~lines are in agreement, within the uncertainties, with those obtained by \cite{wan09} in terms of peak offsets and line flux values.\\

H$\alpha$ + [NII]$\lambda$6550\AA$,\lambda$6585\AA~ emission lines -- As in the case of the [OIII]$\lambda$5008\AA, also the H$\alpha$ emission line shows a clear double-peaked profile (see Fig.~\ref{Halpha_fit}). 
To include both the H$\alpha$ and the [NII] lines in our anlaysis, we 
fitted the residual spectrum of \object{SDSS~J1431+4358} between 6525\AA~and 6615\AA. 
The H$\alpha$ profile requires two Gaussian components, both of which are characterized by  
a  FWHM on the same order as the [OIII]$\lambda$5008\AA~widths. Conversely,  the [NII] emission lines can be fitted well by a single broader Gaussian having FWHM similar to the combination of the two (blue and red)  
[OIII]$\lambda$5008\AA~components (i.e., $\sim$650-700 km s$^{-1}$).
However, we note that the worse resolving power at higher wavelengths (from $\sim$2.5\AA@3800\AA~to$\sim$3.6\AA@9000\AA), combined
with the larger critical electron density for de-excitation of the [NII] transition with respect to the [OIII],  can 
broaden the line profile slightly \citep{ost91}. 
This in turns implies that, with the quality of the current spectra, an intrinsically  double-peaked profile for the [NII] emission line would appear as a single broad line.
To test if  the apparent broadening could be due to the presence of double-peaked lines, we replaced each of these lines with two emission lines. We thus make the approximation that  
the two  [NII] emission lines (at 6550\AA~ and 6585\AA) are both double-peaked as well.  We then assumed that their widths are on the same order, in units of km s$^{-1}$, of those estimated for the other narrow emission lines (
i.e., [OIII] and H$\alpha)$. In particular, we tied the [NII] widths to the [OIII] values. In this case,
two Gaussian components are clearly requested to reproduce each of the [NII] emission lines.
The fitting results are showed in Fig.~\ref{Halpha_fit} while the best-fit parameters  are reported in Table~\ref{sdss15_oiii}.\\

[SII]$\lambda$6718\AA$,\lambda$6733\AA~ emission lines --
The spectral region around the [SII] emission lines is
shown in Fig.~\ref{SII_fit}.
As for the [NII] emission lines, we assumed the emission line widths of the [SII] components to be equal to 
those of the [OIII] emission lines, in units of km s$^{-1}$. Although with larger uncertainties, 
we found that also in this case,  each of the [SII]  emission lines can be well reproduced by the combination of two Gaussian components (see Fig.~\ref{SII_fit} and Table~\ref{sdss15_oiii}).

\begin{table*}[h!]
 \begin{minipage}[t]{1\textwidth}
        \begin{center}
  \caption{Best-fit values of the most prominent emission line components in the SDSS--DR15 spectrum of \object{SDSS~J1431+4358} (see Fig.~\ref{sdss15_spectrum}). The best-fit parameters were obtained by imposing  the widths of all the blue (red) Gaussian components to be on the same order, in units of km~s$^{-1}$, of the width estimated for the blue (red) [OIII]$\lambda$5008.2\AA~component, where 
 FWHM([OIII]$_{\rm blue}$)=288$\pm$42 km~s$^{-1}$ and FWHM([OIII]$_{\rm red}$)=329$\pm$72 km~s$^{-1}$.}
  \label{sdss15_oiii}
 \scalebox{1}{
  \begin{tabular}{ccccccc}
   \hline
 Spectral line & $\lambda_{\rm peak}$ & $\Delta \lambda_{\rm peak}$ &$\Delta {v}_{\rm peak}$   &  $v_{\rm red} - v_{\rm blue}$& Flux & Luminosity     \\
                     &   [$\AA$]                       &    [$\AA$]                                 &       [km/s]                       &       [km/s]                       &        [10$^{-17}$ erg/cm$^{2}$s ]  &   [10$^{40}$ erg/s]\\
         (1)       &      (2)   &   (3)     &     (4)    &   (5)      &   (6)                                      &    (7)    \\            
                     \hline
                     \hline
 ~~~\\
\multirow{ 3}{*}{ $\rm {H\beta}{\lambda 4862.7}$$\AA$}  & 4860.3$\pm$2.0 & $\sim$-2.4 & $\sim$-148 & \multirow{ 3}{*}{$\sim$250} & $\sim$12 & $\sim$0.3\\
 ~~~\\
                                                        & 4864.3$\pm$2.0 & $\sim$1.6  &  $\sim$100 & & $\sim$9 &$\sim$0.2\\
 ~~~\\
 ~~~\\
\multirow{ 3}{*}{ $\rm {[OIII]}{\lambda 4960.3}$$\AA$}  & 4957.1$\pm$1.5 &-3.2$\pm$2.1& -194$\pm$128 & \multirow{ 3}{*}{436$\pm$100} & 13$\pm$10& 0.3$\pm$0.6\\
 ~~~\\
                                                        & 4964.3$\pm$0.7 & 4.0$\pm$1.0   &  242$\pm$60 & &31$\pm$20 &0.7$\pm$0.2\\
 ~~~\\
 ~~~\\
 \multirow{ 3}{*}{ $\rm {[OIII]}{\lambda 5008.2}$$\AA$}  & 5005.6$\pm$0.5 & -2.6$\pm$0.7& -156$\pm$42 & \multirow{ 3}{*}{366$\pm$42} & 77$\pm$15& 1.8$\pm$0.4\\
 ~~~\\
                                                        & 5011.7$\pm$0.5 & 3.5$\pm$0.7   &  210$\pm$42 & &89$\pm$28 &2.1$\pm$0.4\\
 ~~~\\
 ~~~\\
 \multirow{ 3}{*}{ $\rm {[NII]}{\lambda 6549.9}$$\AA$}  & 6545.0$\pm$2.0 & -4.9$\pm$2.8& -223$\pm$130 & \multirow{ 3}{*}{310$\pm$110} & 22$\pm$ 14& 0.5$\pm$0.3\\
 ~~~\\
                                                        &6551.8$\pm$1.3 & 1.9$\pm$1.8   &  87$\pm$84 & &37$\pm$24 &0.9$\pm$0.4\\
 ~~~\\
 ~~~\\
 \multirow{ 3}{*}{ $\rm {H\alpha}{\lambda 6564.6}$$\AA$}  & 6561.5$\pm$0.5 & -3.1$\pm$0.7& -142$\pm$32 & \multirow{ 3}{*}{348$\pm$30} & 87$\pm$ 22& 2.1$\pm$0.4\\
 ~~~\\
                                                        & 6569.1$\pm$0.4 & 4.5$\pm$0.6   &  206$\pm$26 & &94$\pm$24 &2.2$\pm$0.4\\
 ~~~\\
 ~~~\\
 \multirow{ 3}{*}{ $\rm {[NII]}{\lambda 6585.3}$$\AA$}  & 6581.5$\pm$1.3& -3.8$\pm$1.8& -173$\pm$85 & \multirow{ 3}{*}{296$\pm$87} & 96$\pm$ 25& 2.2$\pm$0.5\\
 ~~~\\
                                                        & 6588.0$\pm$1.4 & 2.7$\pm$1.9  &  123$\pm$90 & &110$\pm$30 &2.6$\pm$0.5\\
 ~~~\\
 ~~~\\
  \multirow{ 3}{*}{ $\rm {[SII]}{\lambda 6718.3}$$\AA$}  & 6714.3$\pm$1.2 & -4.0$\pm$1.7& -179$\pm$76 & \multirow{ 3}{*}{335$\pm$120} & 36$\pm$10 & 0.8$\pm$0.2\\
 ~~~\\
                                                        & 6721.8$\pm$2.4 & $\sim$3.5  &  $\sim$156& &32$\pm$16 &0.7$\pm$0.3\\
 ~~~\\
 ~~~\\
 \multirow{ 3}{*}{ $\rm {[SII]}{\lambda 6732.7}$$\AA$}  & 6728.8$\pm$3.4& $\sim$-3.9& $\sim$-174& \multirow{ 3}{*}{$\sim$263} & $\sim$24& $\sim$0.6\\
 ~~~\\
                                                        & 6734.7$\pm$3.4 & $\sim$2.0  & $\sim$89 & &$\sim$40 &$\sim$0.9\\
 ~~~\\
 ~~~\\
\hline
\hline
\end{tabular}
}
\end{center}

{\bf Notes.}  Col. (1): Spectral lines (rest-frame vacuum wavelengths). 
Col. (2): Rest-frame wavelengths of the peaks of the blue- and redshifted emission line components.
Col. (3): Doppler shifts of the blue and red emission line components. 
Col (4): Line-of-sight velocity offsets of the blue- and redshifted components. 
Col. (5): Line-of-sight velocity offset between red and blue peaks.  
Col. (6)-(7): Fluxes and luminosities, corrected for Galactic extinction, of the blue- and redshifted components. \\
  
\end{minipage}
\end{table*}

\begin{figure*}[h!]
        \includegraphics[scale=0.7]{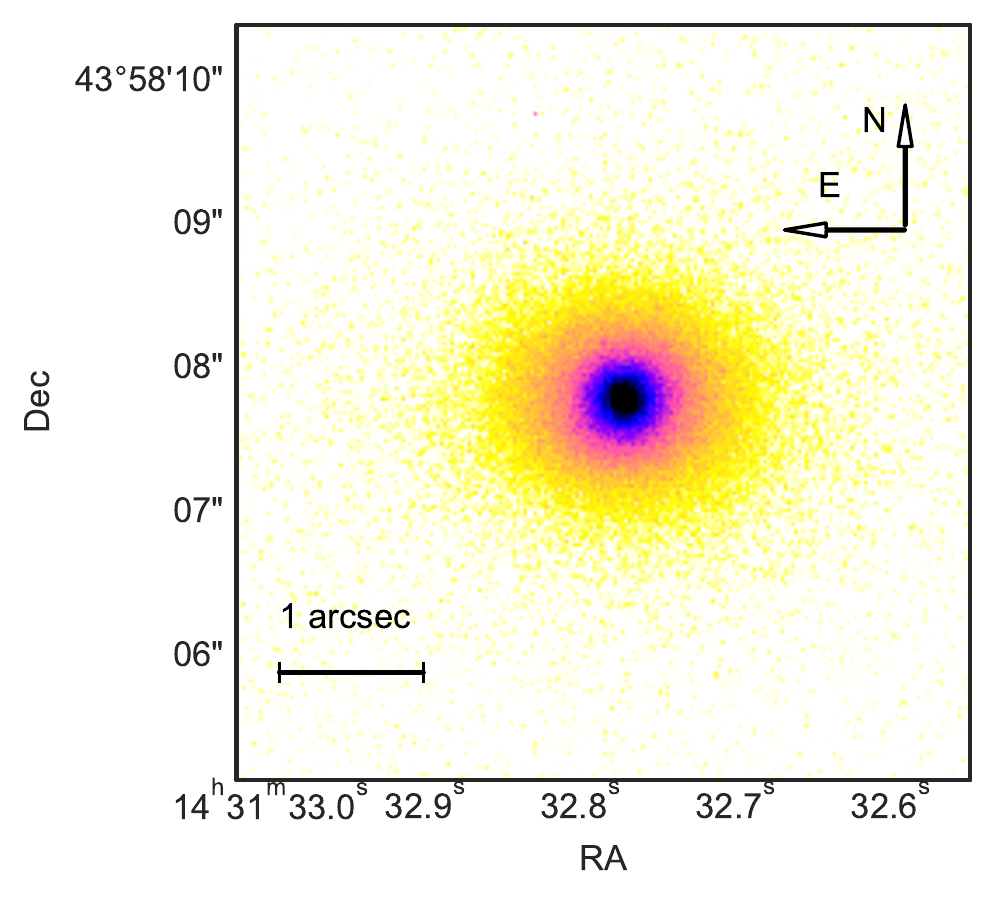}
        \hskip 1truecm
        \includegraphics[scale=0.7]{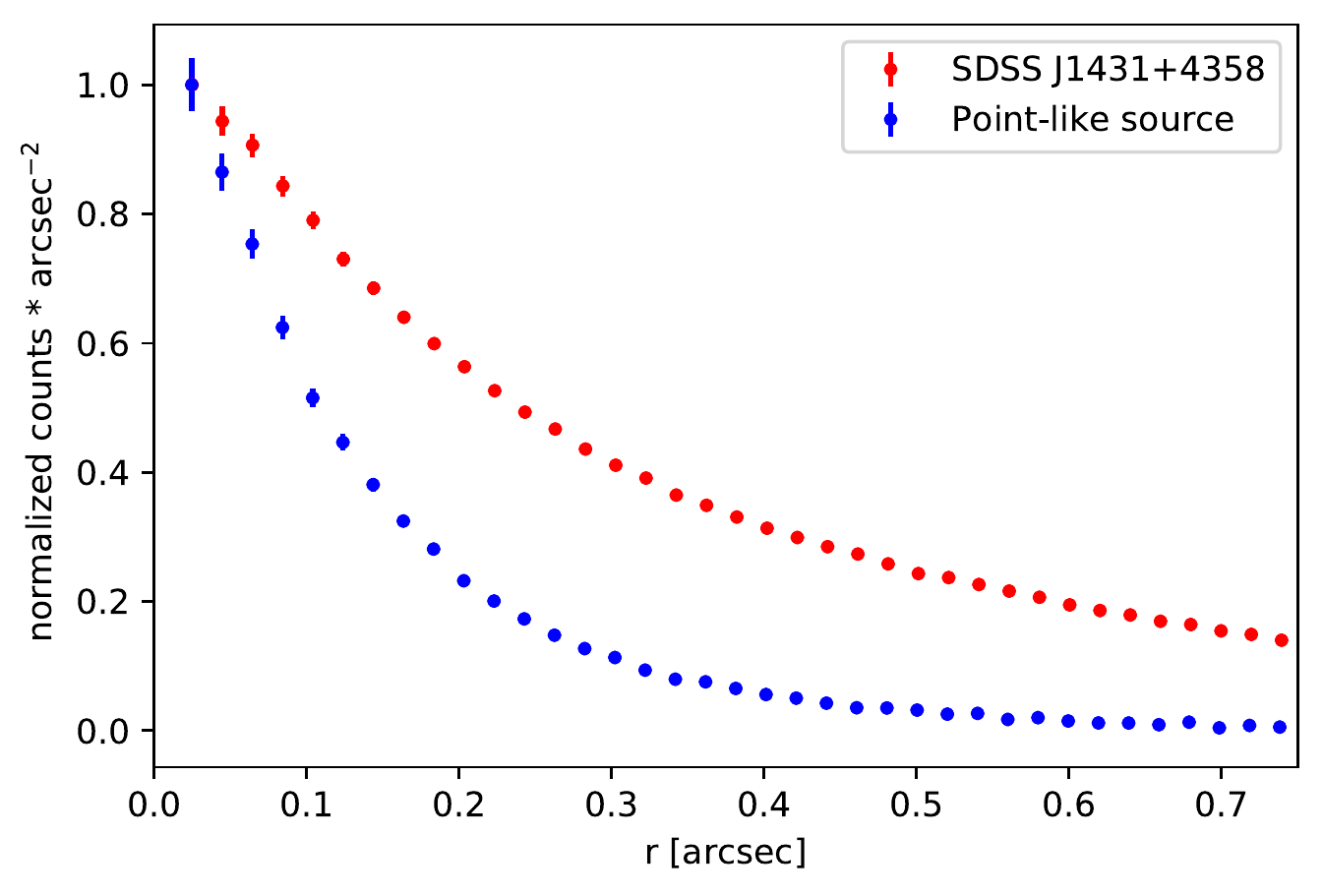}
    \caption{Left panel: Zoomed (5$\arcsec$$\times$5$\arcsec$) K-band image of \object{SDSS~J1431+4358} taken with LBT+LUCI+FLAO in diffraction-limited mode. Right panel: Normalized radial profiles in the K-band,  measured within circular annuli centered on the target (\object{SDSS~J1431+4358}, red points) and on the point-like source (blue points)  used for the PSF reconstruction  at about  5.5$\arcsec$ from the target (see Sect.\ref{lbt}).}
    \label{k_image}
\end{figure*}

\subsection{Observations by LBT}
\label{lbt}

We observed the dual AGN candidate, \object{SDSS~J1431+4358} (K$_{\rm S}$$\sim$13.8 mag\footnote{The K$_{\rm S}$ magnitude was taken by the final release of the  
2MASS extended objects, https://irsa.ipac.caltech.edu/cgi-bin/Gator/nph-dd.}
)  with the  LBT-LUCI2 camera  \citep[one of the two LBT Utility Cameras in the Infrared;][]{sei03} with the First Light Adaptive Optics system \citep[FLAO;][]{esp12} in diffraction limited mode. The observations were performed on 9 May 2018 (Program ID: LBT-2018A-C1349-12, P.I. Severgnini) in the K band using the N30-camera (pixel scale of 0.015 arcsec/pixel) for a
total exposure time of 1800 sec and on-source exposure time of 1200~sec. 
The observations were conducted under a natural seeing condition of 0.7" (FWHM), which allowed us to reach an image quality of 0.15$\arcsec$ (FWHM).
 
The LUCI data reduction pipeline developed at INAF -- Osservatorio Astronomico di Roma{\footnote{http://www.oa-roma.inaf.it}} was used to perform the basic data reduction that includes dark subtraction, bad pixel masking, cosmic ray removal, flat fielding, and sky subtraction. Astrometric solutions for individual frames were obtained, and the single frames were then recombined using a weighted co-addition 
to obtain the final stacked image normalized to 1 sec of exposure. 
In Fig.~\ref{k_image} (left panel), we show a zoom on the 5$\arcsec$$\times$5$\arcsec$ reduced K-band image. The target is clearly resolved  (Fig.~\ref{k_image}, right panel) and bulge dominated. 
As expected on the basis of the optical spectrum (see Fig.~\ref{sdss15_spectrum}) and on the basis of the R-K color ($\sim$2.3 mag), typical of local normal galaxies \citep[e.g.,][]{man01,cha06}, the host galaxy emission not only dominates the optical bands but also the NIR band.
This implies that, if a nuclear component is present, 
regardless of whether it is a
single or a double component, its infrared emission is
completely diluted by the host galaxy emission and 
therefore, it cannot be directly detected without a proper bulge component subtraction from the NIR image.
\vskip 0.1truecm
\paragraph{Photometric fitting residuals: GALFIT}
~\\

\noindent Since the double nuclei could be hidden by the galaxy
continuum emission, we then proceeded to fit the galaxy profile and subtracted it from the data.  We used 
the 2-D fitting algorithm GALFIT  \citep{pen02,pen10} and adopted for the 
 galaxy  the 1--D  S{\'e}rsic
profile \citep{ser63}
\begin{center}
\begin{equation}
I(r)=I_{\rm e} \exp \left[-b_{\rm n}\left(\left(\frac{r}{r_{\rm e}}\right)^{\frac{1}{n}} - 1\right)\right]
,\end{equation}
\end{center}
where  $I_{\rm e}$ is the pixel surface brightness at radius $r_{\rm e}$ and the shape of the profile is determined by the S{\'e}rsic index $n$. This function allows us to model 
 an exponential disk  ($n$=1) or a classical bulge component,
which is usually characterized by  $n>$2, and in particular the de Vacouleurs
profile  \citep{dev48} with $n$=4. 

To fit the galaxy profile of \object{SDSS~J1431+4358}, we convolved
a  2--D S{\'e}rsic model  with the point spread function (PSF).
For the PSF, we used the true image of a high S/N point-like source
in the LUCI field of view, located at 5.5$\arcsec$ from the target galaxy
(the 1-D profile is shown in the right panel of Fig.~\ref{k_image}, blue points).
In the fitting procedure we also used a flat background sky and allowed  the 
 centroid (x,y),  the S{\'e}rsic index n,  the effective radius $R_{\rm e}$ and the
total magnitude to vary.
The galaxy contribution  is  well reproduced ($\chi^2$/$d.o.f$ = 1.03)
with a S{\'e}rsic index n=4.8$\pm$0.1 and an effective radius $R_{\rm e}$=(2.7$\pm$0.1)$\arcsec$=(4.8$\pm$0.1) kpc.

\begin{figure}[t!]
\centering
        \includegraphics[scale=0.6]{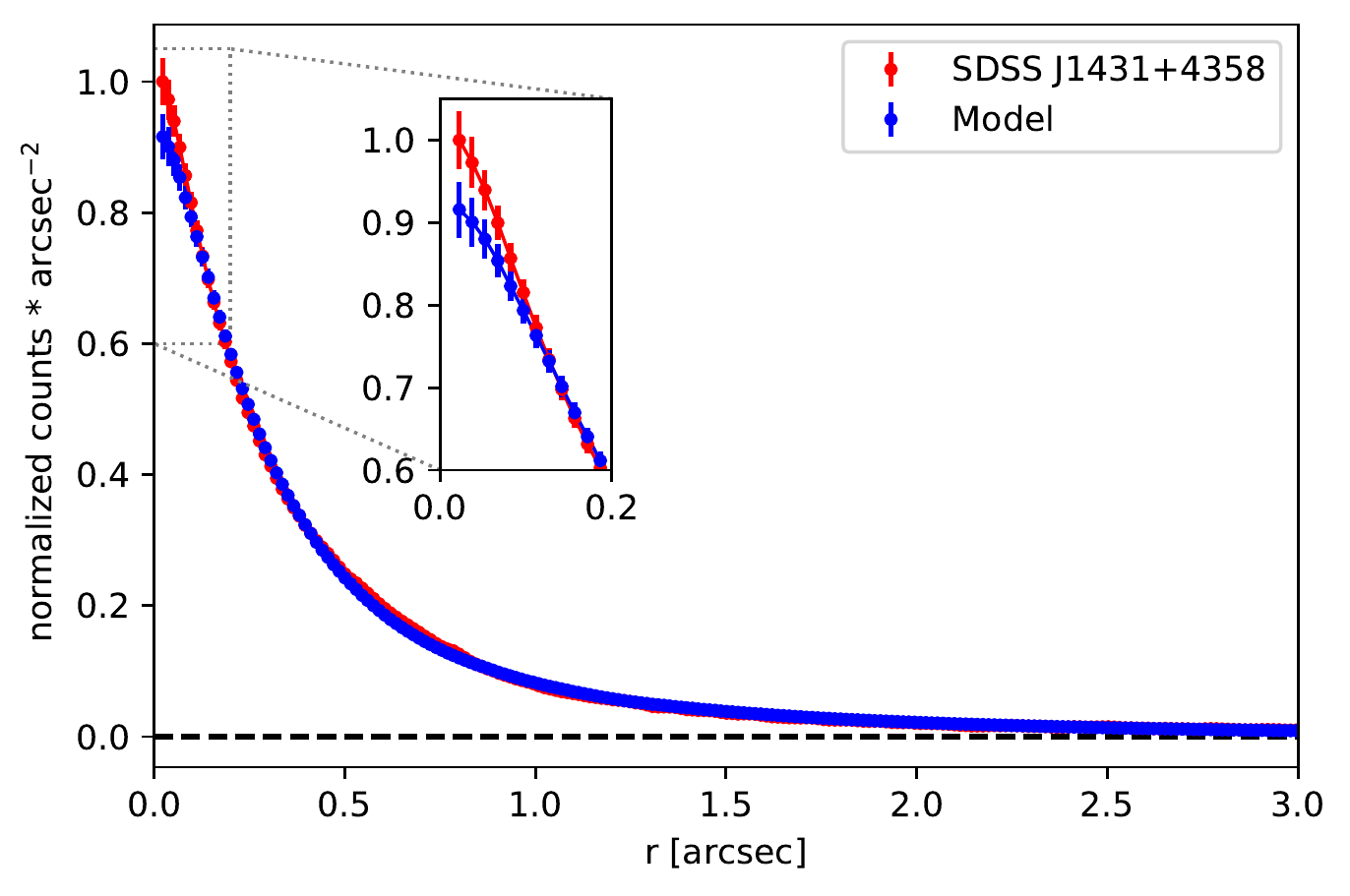}
    \caption{Radial profiles, as  measured within circular annulus on the K-band image and centered on the target (\object{SDSS~J1431+4358}, red points), compared with the surface brightness from the best-fitting   S{\'e}rsic  model profile (blue points). Both profiles are normalized to the peak value of the source.}
    \label{galaxy_model_profile}
\end{figure}

\begin{figure*}[t!]
\centering
        \includegraphics[scale=0.7]{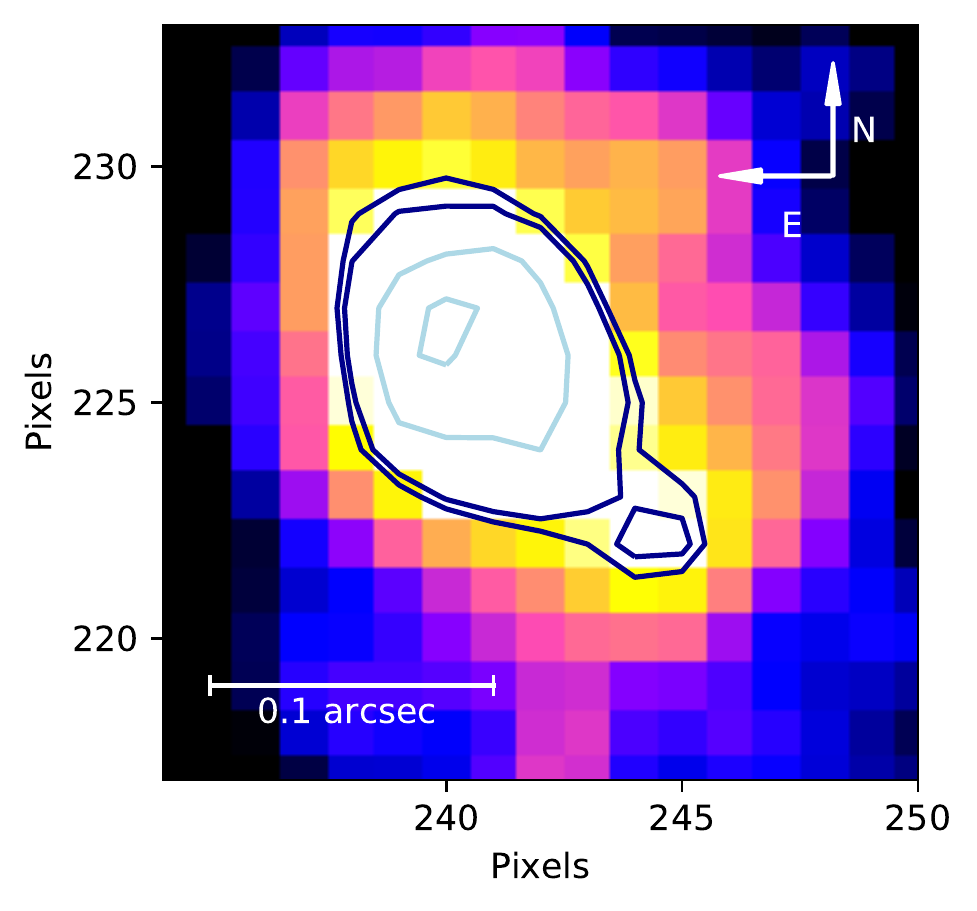}
             \includegraphics[scale=0.7]{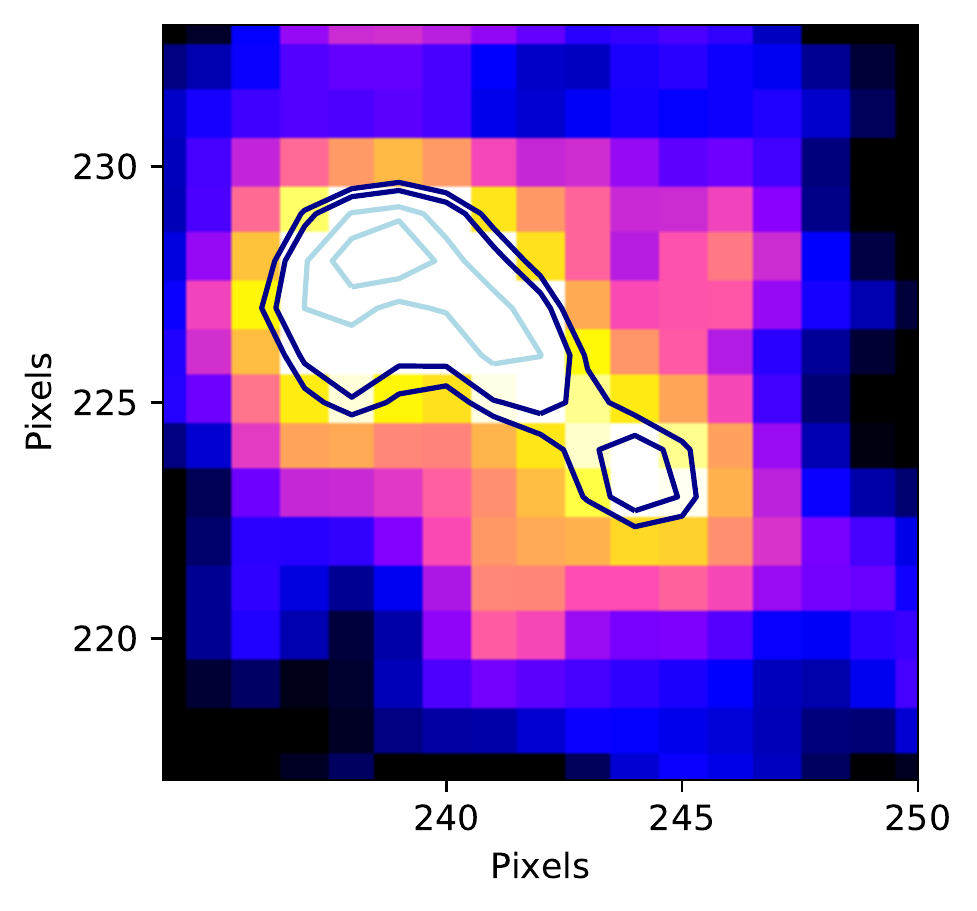}
    \caption{ Left panel: K-band residual image (0.24$\arcsec$$\times$0.24$\arcsec$)  centered on \object{SDSS~J1431+4358}. The image was obtained by smoothing the real K-band residual image with a
Gaussian filter
($\sigma$=1 pixel=0.015$\arcsec$; see Sect.~\ref{lbt}). Contour plots are overlaid: The dark blue (light blue) lines indicate the pixels with a value of 2.5 and 2.7 (3 and 4) times 
the standard deviation of the background of the unsmoothed residual image. Right panel: Simulated image, obtained by considering two point-like sources (relative distance of 0.12$\arcsec$ and with K-magnitudes equal to 19.5 mag and  20.3 mag), the true PSF image, and the background sky (see Sect.~\ref{lbt}  for details). The image was obtained by smoothing the simulated image with a Gaussian filter
($\sigma$=1 pixel=0.015$\arcsec$). Contour plots are overlaid: The dark blue (light blue) lines indicate the pixels with a value of 2.5 and 2.7 (3 and 4) times 
the standard deviation of the background of the unsmoothed simulated image. }
    \label{contours}
\end{figure*}

In Fig.~\ref{galaxy_model_profile}, we compare the radial profile measured  from the K-band image, using  circular annuli centered on the nuclear source (red points), and the predicted radial profile derived from the best-fitting S{\'e}rsic model (blue points).
Within $\sim$0.2$\arcsec$ of radius, positive residuals  with respect to the S{\'e}rsic model are present, possibly resulting from the AGN contribution. The residual image produced by GALFIT in the central 0.24$\arcsec$x0.24$\arcsec$ region  is shown in Fig.~\ref{contours} (left panel). 
The image reveals the presence of  two possible distinct nuclear components at projected separation of $\sim$0.12$\arcsec$, as measured from the distance of the two brightest pixels. The northern and brighter component has a projected position almost spatially coincident with
the host galaxy center. The measured positional angle is $\approx$50 deg, in agreement with the orientation measured for the NLR 
\citep[][$PA_{\rm [OIII]} \sim$59 $\deg$]{nev16}. 
The contour levels in Fig.~\ref{contours} (left panel)  
show the pixels with a value from 2.5 to 4 times the standard deviation of the background.

In order to assess whether two point--like sources at such a small angular separation can be resolved by AO assisted LUCI observations, we performed some basic simulations with GALFIT.
We modeled the two point-like sources by convolving two Gaussians, which have the minimum FWHM value allowed by GALFIT (FWHM=1.5$\times$10$^{-3}$ arcsec), with the true PSF of the image (see above). 
We then added the true background r.m.s. derived by  the real image.
We produced several simulated images for different values of the relative intensity and separation between the two point-like sources. 
We verified that down to an angular separation of about 0.12$\arcsec$ the two simulated sources appear to be resolved. 
  In particular, we found that the two observed NIR blobs can be reproduced well by two  sources at $\sim$0.12-0.2$\arcsec$ apart (corresponding to $\sim$215-360 pc). The K-band magnitudes of the northern and southern simulated
sources are $\sim$19 mag and $\sim$20 mag, respectively.
Since these magnitudes were estimated by imposing that the emission peaks in the simulated images had the same value as the peaks in the real image, they strongly depend on 
the adopted  background noise. Using different background r.m.s. image derived by different zones of the real image, we estimated that the uncertainties on the magnitudes are on the order of 1 mag.
As an example of our simulations, in Fig.~\ref{contours} (right panel) we show one of the 
simulated residual maps obtained by using two point-like sources with  a relative distance of 0.12$\arcsec$ and with K-magnitudes equal to 19.5 mag and  20.3 mag. 

\subsection{Swift-XRT observation}
\label{xrt}
As X-ray data were not available for \object{SDSS~J1431+4358}, we recently were awarded an explorative observation  with the  {\it Swift}-XRT \citep[][]{bur05}. This observation was carried out  through a  Target  of  Opportunity  (ToO)  request (5 ksec, Target ID: 13292, observation date: 2020 March 18). The 0.3-10 keV image was extracted  using the online XRT data product generator{\footnote{https://www.swift.ac.uk/user\_objects/}} \citep{eva07,eva09}. 
\object{SDSS~J1431+4358} was not detected down to a 3$\sigma$ limiting flux of about 10$^{-13}$ erg cm$^{-2}$ s$^{-1}$. We note that the source was not detected in the 0.3-4 keV and 4-10 keV images either.

\section{Results}
\label{results}

Our analysis of the SDSS spectrum (Sect.~\ref{optical}) confirms the type 2 spectroscopic classification of \object{SDSS~J1431+4358} \citep{wan09,ge12}.
The lack  of significant broad emission line components up to the H$\alpha$ wavelengths suggests  
an intrinsic A$_{\rm V}$$\geq$3 mag \citep[e.g.,][]{gil01}. 

We found that both the
[OIII]$\lambda$5008\AA~and the H$\alpha$ emission lines show a clear double-peaked profile, which
is well fitted by two narrow Gaussian components.
Once we constrain the FWHM of the other emission lines, which are present 
 in the spectrum, to be on the same order of  the width of the [OIII] lines, we also found that the [NII] and the [SII] emissions could be accounted for by two  Gaussian emission line components. 
The shifts  between the blue (red) peak and the systemic redshift are similar, within the uncertainties,  for all the observed transitions across the full spectral range (see Table~\ref{sdss15_oiii}).
In particular, for the [OIII]$\lambda$5008\AA~and the H$\alpha$ emission lines, for which the presence of double components are clearly evident, neither the blue  nor the red  peaks are consistent with the nominal systemic velocity of the host galaxy (derived independently from the stellar absorption lines).
We found that the flux ratios between the blue and red peaks are on the order of one for most of the more prominent emission lines.

Regarding the H$\beta$ transition, a weak narrow emission line is visible only in the residual spectrum, while in the observed spectrum the relevant stellar absorption component fully overwhelms it.
Owing to the low S/N, the emission line profile can be fitted well by either a single 
(see  Sect.~\ref{optical}) or 
 double Gaussian components (see Table~\ref{sdss15_oiii}). 

\begin{figure}[t!]
\centering
        \includegraphics[scale=0.6]{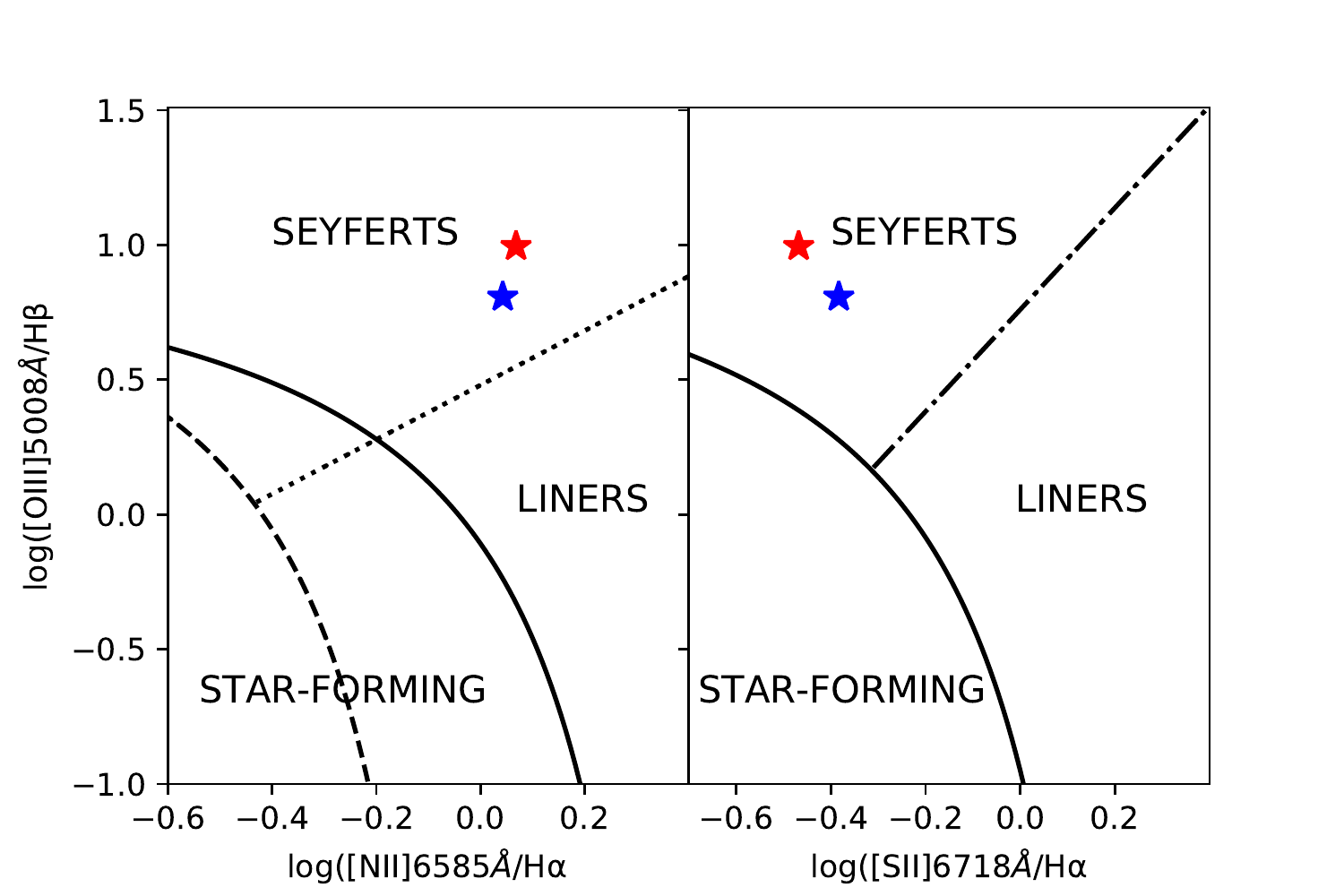}
    \caption{BPT diagram of the blue (blue star symbol) and red (red star symbol) 
    components detected in the optical spectrum of \object{SDSS~J1431+4358}. 
    The solid and dashed lines divide the regions mainly populated by star-forming galaxies and         AGN  from \cite{kew01,kew06} and \cite{kau03}, respectively.      
The dotted and dot-dashed  lines demarcate the
    Seyfert and LINER zones from \cite{cid10} and \cite{kew06}, respectively.}
    \label{bpt}
\end{figure}

 We then investigated the nature of
the ionizing source producing the blue and red peak components through the BPT 
diagnostic diagram (see Fig.~\ref{bpt}).
From this figure we can see that both the blue and red peaks detected in the \object{SDSS~J1431+4358} spectrum 
occupy the AGN locus, where the gas is mainly  photoionized by the AGN emission  \cite[see also][]{ge12}.

In agreement with the results found by \cite{nev16}, our analysis  does not rule out the presence of an ionized outflow component; indeed, a faint [OIII] broader (FWHM$\sim$500 km s$^{-1}$) blue wing is  tentatively detected. However, this component cannot explain the
double-peaked profile of the [OIII] line.
This latter 
is reproduced well by two further narrow Gaussian components (blue- and redshifted, respectively) with similar peak intensities.

The K-band emission of \object{SDSS~J1431+4358}, as well as the optical emission, is fully dominated by the host galaxy contribution.
Thanks to the tight sub-arcsec sampling provided by the LBT-LUCI NIR camera in AO mode (0.015 \arcsec/pixel), we unveiled  significant residuals in the nuclear zone (within 0.2 arcsec) with respect to the expectation for a single galaxy model.
After subtracting the host
galaxy contribution, two spatially distinct NIR components
emerge at sub-kiloparsec separation.
Intriguingly, the positional angle of the two NIR
sources ($PA_{\rm NIR} \sim$50 $\deg$) is close, within a typical error of 15 $\deg$, to that estimated for the two emitting  NLR  \citep[$PA_{\rm [OIII]} \sim$59 $\deg$;][]{nev16}. However, the authors do
not provide the physical projected offset of the two NLR centroids.

\object{SDSS~J1431+4358} was not detected in our  {\it Swift}-XRT observation. The 3$\sigma$ upper limit on the X-ray flux  of about 10$^{-13}$ erg cm$^{-2}$ s$^{-1}$ (see Sect.~\ref{xrt}) could be an indication of the presence of high obscuration in  the X-ray band as well.
We calculated the most plausible unabsorbed X-flux from the  [OIII] de-reddened luminosity ($L^c_{\rm [OIII]}$). We first applied the  
hydrogen Balmer decrement correction to the total luminosity\footnote{We considered as total line luminosity the sum of the luminosities of the blue and red components reported in Table~\ref{sdss15_oiii}.} of the [OIII]$\lambda$5008\AA~emission line,\begin{equation}
L^c_{\rm [OIII]}=L_{\rm [OIII]}{\left[\frac
{(H\alpha/H\beta)_{obs}}{3.0}\right]}^{2.94}
\end{equation}
where $(H\alpha$/$H\beta)_{\rm obs}$ is the observed line ratio \citep{ost06}.
Then, by adopting the $L_{\rm X}$-$L^c_{\rm [OIII]}$ relation reported by \citet{lam09} and by \citet{pan06}, we predicted an unabsorbed 
 0.3-10 keV flux on the order of about 5$\times$10$^{-13}$ erg cm$^{-2}$ s$^{-1}$ and 
10$^{-12}$ erg cm$^{-2}$ s$^{-1}$, respectively (well above the derived upper limit).
We then assumed that the intrinsic X-ray emission  is a power law with the typical photon index of  $\Gamma$=1.8
and estimated the amount of absorption that could explain the nondetection in the XRT observation.
We found that we need a column density of the absorbing gas in excess of 10$^{22}$--10$^{23}$ cm$^{-2}$.

\section{Discussion}
\label{discussion}

\object{SDSS~J1431+4358} was first selected as a type 2 dual AGN candidate  by \cite{wan09} 
on the basis of the double-peaked profile of the [OIII]$\lambda$5008\AA~emission line. 
The presence of  narrow emission lines with double-peaked profiles  was 
subsequently confirmed by  \cite{ge12}, who pointed out that both the blue- and the redshifted emission lines are  produced by AGN photoionization. Our independent analysis corroborates these results.

Interestingly, the misalignment measured by \cite{nev16} between the position of the NLR and the major axis of the host galaxy ($PA_{\rm [OIII]}$=59 $\deg$ and $PA_{\rm gal}$=87 $\deg$)  disfavors the  possibility that double-peaked NLRs are tracing the gas kinematics of the host galaxy.

In these previous works \citep[][]{wan09,ge12,nev16}, the lack of a dual core detection in the SDSS images left equally open  the hypotheses of a single SMBH or of
double SMBHs. In the first case, the kinematical origin of the detected double-peaked profiles 
could be due to outflows or to rotational effects of a single NLR disk-like, while in the second case, they could trace the orbital
motion of the two SMBHs around a common central potential.

The typical signature for outflowing and/or inflowing components in integrated galaxy spectra is the presence of fainter and blue and/or red broader emission lines superimposed on the narrow emission line centered at the systemic velocity \citep[e.g.,][]{har14,har16}.
The spectroscopical analysis  presented in this work confirms the possible presence of an ionized outflow component, previously also detected by \cite{nev16}. However, we found that the same outflow mechanism can hardly explain the presence of  two narrow emission lines
(blue- and redshifted, respectively) with similar peak intensities (as in the case of \object{SDSS~J1431+4358}, see Table~\ref{sdss15_oiii}).
The receding sides of ionized outflows traced by optical nebular lines are usually undetected (or extremely faint) owing to extinction by dust in the center of the galaxy and in the outflow itself. So, in the case of \object{SDSS~J1431+4358}, there are two possibilities. First, if we ascribe only the blue component to an outflow (in addition to the blueshifted wing at $\sim$-700 km s$^{-1}$ tentatively detected in the [OIII] transitions and shown in Fig.~\ref{OIII_fit}), then it would mean that the red component of the line traces the systemic NLR emission, which would be redshifted by $\sim$210 km s$^{-1}$ with respect to the host galaxy redshift determined via the stellar absorption. Second, if instead we ascribe both components to a symmetric red- plus blueshifted outflow, then it would mean that there is no dust obscuration since both components are detected at the same level, and that there is no NLR emission at the systemic velocity. Basically, it would imply that all of the nebular emission comes from an outflow. This would be very odd, and it is a situation that has not been observed even in the most powerful outflows observed so far; there is always a quiescent narrow line component. So the only viable interpretation is that neither of the two components traces an outflow.

According to \cite{smi12}, a flux ratio on the order of one between the blue and red peaks 
could be considered as an indication for the presence of a single, rather than a double, ionizing source.
However, we note that, while in the majority of the cases this could be true, it cannot  be used to discard  the presence of two distinct SMBHs in individual systems. The hydrodynamic simulations of galaxy mergers presented by \cite{ble13} showed that in the presence of dual SMBHs, emission line profiles with almost even peaks may arise at 
one viewing angle at least.
The same authors
concluded that  AGNs with uneven-peaked narrow emission line profiles have at most a modestly higher probability of containing two SMBHs than AGNs
with an even-peaked profiles.

Because of the large amount of nuclear absorption affecting the AGN intrinsic emission (i.e., the power law plus the broad emission line components), the source appears as a bulge-dominated galaxy up to the NIR band. After subtracting the host galaxy contribution 
in the high-resolution NIR LBT image, we detect, for the first time, two compact components at a  sub-kiloparsec projected separation in the \object{SDSS~J1431+4358} core.
While the NIR imaging data alone do not allow us to discard the possibilities that the fainter NIR core can be associated with a smaller merging stellar bulge or to an outflow component,  the close spatial alignment between the two NIR cores and the two regions where double-peaked AGN  emission lines originate \citep{nev16} supports the presence of a dual AGN.

Under this hypothesis, the minimum  estimated distance of the two SMBHs is about 215-360 pc (see Sec~\ref{lbt}). This implies that the two putative accretion disks (of typical radius less than 0.01 pc) should be undisturbed and could illuminate the surrounding gas forming two NLRs. 
Thus, the double-peaked profiles could trace the motions of  two distinct or semi-detached NLRs orbiting around a common central potential. 

An alternative scenario, in which double peaks are produced  by two almost co-spatial and disturbed
NLRs forming a single envelope rotating around the two SMBHs cannot be fully discarded, but it is less likely. In this case, the narrow  emission line regions may be partially destroyed or deformed. A more complex emission line profiles would be expected with respect to the almost regular double profiles observed for \object{SDSS~J1431+4358}   \citep[see, e.g.,][]{pop12}

A further possibility is that we are seeing the rotational effect associated only with the brightest and dominant of the two NLRs, that is, the NLR that is rotating around the most powerful  of the two SMBHs \citep[][]{pop12,ble13}.

Only future high-resolution and spatially resolved spectroscopical observations will provide a definitive answer on the nature  of \object{SDSS~J1431+4358} and on the origin of its double-peaked emission lines.
This will allow us to confirm or rule out the spatial coincidence of the double line emission components with the two NIR cores.

\section{Summary and conclusions}
\label{summary}
We presented the spectral analysis of the archival SDSS-DR15 optical spectrum and  the analysis of new  LBT-LUCI NIR sub-arcsec observation for \object{SDSS~J1431+4358}, a local radio-quiet type 2 AGN previously selected as double AGN candidate. 

As for the optical spectrum, all the most prominent emission lines are characterized by a double-peaked profile. On the basis of the flux line ratios, we found that all the emission line components are mainly produced by AGN photoionization.
Although the presence of outflowing ionized material cannot be disfavored, because a third blue and broad [OIII] emission line is tentatively detected, it can hardly be the cause of the two relatively narrow components of the double-peaked profiles.
Rotational effects due to a single NLR associated with a single SMBH or due to the presence of two SMBHs are the most likely hypothesis for the
double-peaked emission lines observed in our target. 
 
The presence of a dual AGN is strengthened by the high-spatial resolution NIR imaging. 
In particular, the tight sub-arcsec sampling provided by the LUCI-AO camera allowed us to detect  significant residuals in the central zone of the galaxy. Once we subtracted the host contribution, two faint central NIR sources emerged.
Interestingly, the two central cores have a  positional angle similar to 
the positional angle  of the two emitting regions where the [OIII] double peaks most likely originate.
The cause of the apparent faintness in the optical and in the NIR bands of the two putative nuclear sources 
could be explained with the presence  of a large amount of intrinsic obscuration affecting the central region.

Although only  high-resolution and spatially resolved spectroscopy
will definitely set the spatial coincidence of the double NLR regions with the double NIR cores, the multiwavelengths results presented in this work favor the hypothesis of a double sub-kiloparsec scale SMBH. Up to now,  dedicated spatially resolved spectroscopy aimed at confirming the presence of a double AGN  were carried out only on candidates presenting two cores directly discerned in radio, optical or NIR high-resolution imaging \cite[see, e.g.,][and references therein]{fu12,mcg15,mul15,das18}. Only a handful of dual AGNs with a separation smaller than about 0.5 kpc have been discovered so far.  However, the
pre-selection of interesting candidates was generally  performed without subtracting the contribution of the host galaxy contribution.
The analysis presented in this paper shows that the lack of an appropriate host galaxy 
subtraction could prevent us from unveiling the presence of two distinct cores located in a single merged galaxy.  This clearly affects the estimates of their fraction among the AGN population.
Thus, in addition to the discovery of a new promising sub-kiloparsec scale dual AGN, our analysis  also suggests that the real fraction of dual AGNs among double-peaked emission line sources could be underestimated
\citep{liu10, com12, fu12, mul15,mcg15}.

\begin{acknowledgements}
We thank the anonymous referee for her/his valuable suggestions which improved the quality of the paper. We acknowledge the support from the LBT-Italian Coordination Facility for the
execution of observations, data distribution and reduction. VB, RDC, RS, PS and AZ acknowledge financial contribution from
the agreements ASI-INAF n.2017-14-H.0 and n.I/037/12/0.
This work is based on data supplied by the UK Swift Science Data Centre at the University of Leicester. The SDSS is managed by the Astrophysical Research Consortium for the Participating Institutions of the SDSS Collaboration (see http://www.sdss.org/collaboration/citing-sdss/).

\end{acknowledgements}

%
   \bibliographystyle{aa} 
   \bibliography{aa_severgnini_final} 
%

\end{document}